\documentstyle[prd,aps,graphics]{revtex}

\begin{document}
\def\la{\mathrel{\mathpalette\fun <}}
\def\ga{\mathrel{\mathpalette\fun >}}
\def\fun#1#2{\lower3.6pt\vbox{\baselineskip0pt\lineskip.9pt
\ialign{$\mathsurround=0pt#1\hfil##\hfil$\crcr#2\crcr\sim\crcr}}}  
\def\lrang#1{\left\langle#1\right\rangle}

\begin{center}
{\Large \bfseries High-mass dimuon, $B \overline{B}$ and $B \rightarrow J/\psi$ 
production in ultrarelativistic nucleus-nucleus interactions~\footnote{Talk given at 
International Workshop on Physics of the Quark-Gluon Plasma, Palaiseau, France, 
September 4-7, 2001}.}  

\vskip 3mm 

I.P.Lokhtin

\vskip 2mm

{\small{\it M.V.Lomonosov Moscow State University, D.V.Skobeltsyn Institute of Nuclear 
Physics}} 
\end{center}

\vskip 3mm

\begin{center}
\begin{minipage}{150mm}
\centerline{\bf Abstract} 
The potential of $B$-physics in ultrarelativistic heavy ion collisions is
discussed. The different mechanisms of heavy quark production at high
energies are considered. We analyze the sensitivity of high-mass $\mu^+\mu^-$ pairs 
from $B \overline{B}$ semileptonic decays and secondary $J/\psi$'s from single 
$B$ decays to the medium-induced bottom quark energy loss at LHC energies.  
\end{minipage}
\end{center}

\vskip 5mm

\section {Introduction}   

The production of heavy $b$- and $c$-quarks in hadronic interactions at very high
energies allows one to study the dynamics of hard processes describing the standard as
well as new QCD-physics\cite{ander83,norrbin}. This production can be described
completely in the frame of perturbative QCD under condition the mass of quark $M_Q$ is
much larger than characteristic QCD confinement scale, $M_Q \gg \Lambda_{QCD}$.
The specific interest for heavy quark production in ultrarelativistic nuclear
interactions is due to the intriguing possibility to study behaviour of massive colour 
charge in super-dense QCD-matter -- quark-gluon plasma (QGP), the search and 
investigation of properties of which being one of the goals of modern high energy 
physics (see, for example, reviews~\cite{satz182,muller96,lok_rev,bass_rev}). 
Experimental data obtained at CERN-SPS and RHIC-BNL can be interpreted as a result of 
QGP formation in heavy ion collisions, although alternative explanations can not be 
fully dismissed~\cite{qm01}. It is expected that the most central heavy ion collisions 
at LHC collider might produce practically "ideal" quark-gluon plasma at extremely 
high energy density up to $\varepsilon _0 \sim 0.5$ TeV$/$fm$^3 \gg \varepsilon_{crit} 
\sim 1 $ GeV$/$fm$^3$~\cite{esk}. The inclusive cross section for $b$-quark production 
at LHC ($\sqrt{s}=5.5 A$ TeV for Pb$-$Pb) will be large enough for systematical studies 
of different aspects of $B$-physics, while at RHIC ($\sqrt{s}=200 A$ GeV for Au$-$Au)
the copious production only of $c$-quarks can be expected. The heavy 
quark pairs, created at the very beginning of the collision process, propagate through 
the dense matter and interact strongly with constituents of the medium. 
In-medium gluon radiation and collisional energy loss (see review~\cite{loss_rev} and 
references therein) of heavy quarks can result in experimentally observed modification 
of high mass dilepton spectra~\cite{shur97,kamp98,lin98,lokhtin2}. 
It was also predicted recently that the finite quark mass effects can 
lead to a relative suppression of medium-induced radiation of heavy quarks with 
corresponding enhancement of heavy-to-light $D(B)/\pi$ ratio~\cite{kharz}. 
Since the parton rescattering intensity strongly increases with temperature, formation 
of a "hot" QGP at initial temperatures up to $T_0 \sim 1$ GeV at LHC~\cite{esk} should 
result in much larger parton energy loss compared to "cold" nuclear matter or a 
hadronic gas. 

\section{Mechanisms of heavy quark production and dimuon spectra}

The diagrams for different mechanisms of heavy quark production in high energy 
hadronic interactions are presented in figure 1. Following the classification and 
terminology of paper~\cite{norrbin}, we can distinguish three classes of such
processes with different number of heavy quarks in hard sub-processes vertices 
($2$, $1$ or $0$): "pair creation" (leading order diagrams are shown on the top of 
fig.1); "flavour excitation" (one heavy quark is produced in vertex of hard process
and another quark is created from initial state parton shower); "gluon splitting" 
(both heavy quarks are produced from final state parton shower). 
At present accelerator energies (including RHIC) the bulk of heavy quarks are produced 
due to direct hard scattering ("pair creation"). However, the prediction for LHC 
is that the contribution of gluon splittings in initial- or final-state shower 
evolution to heavy flavour yield can be a significant, about $90\%$
for the whole kinematical range~\cite{norrbin}. 

Since heavy quark pairs are produced at the very beginning of the nuclear collisions,
its propagate through the dense medium. They finally form $B$ and $D$ mesons 
by ``capturing" $u$, $d$ or $s$ quarks during the 
hadronization stage. These mesons will decay with
the average meson lifetimes $c\tau_{B^{\pm}} = 496$ $\mu$m, 
$c\tau_{B^0} = 464$ $\mu$m, $c\tau_{D^{\pm}} = 315$ $\mu$m 
and $c\tau_{D^0} = 124$ $\mu$m.  We note that
$\approx 20 \%$ of $B$ mesons and $\approx 12 \%$ 
of $D$ mesons decay to muons. About 
half of the muons from $B$ decays are produced through an intermediate 
$D$ and contribute to the softer part of the $p_T$-spectrum. 

\begin{figure}[htb]
\begin{minipage}[t]{77mm}
\resizebox{75mm}{75mm}  
{\includegraphics{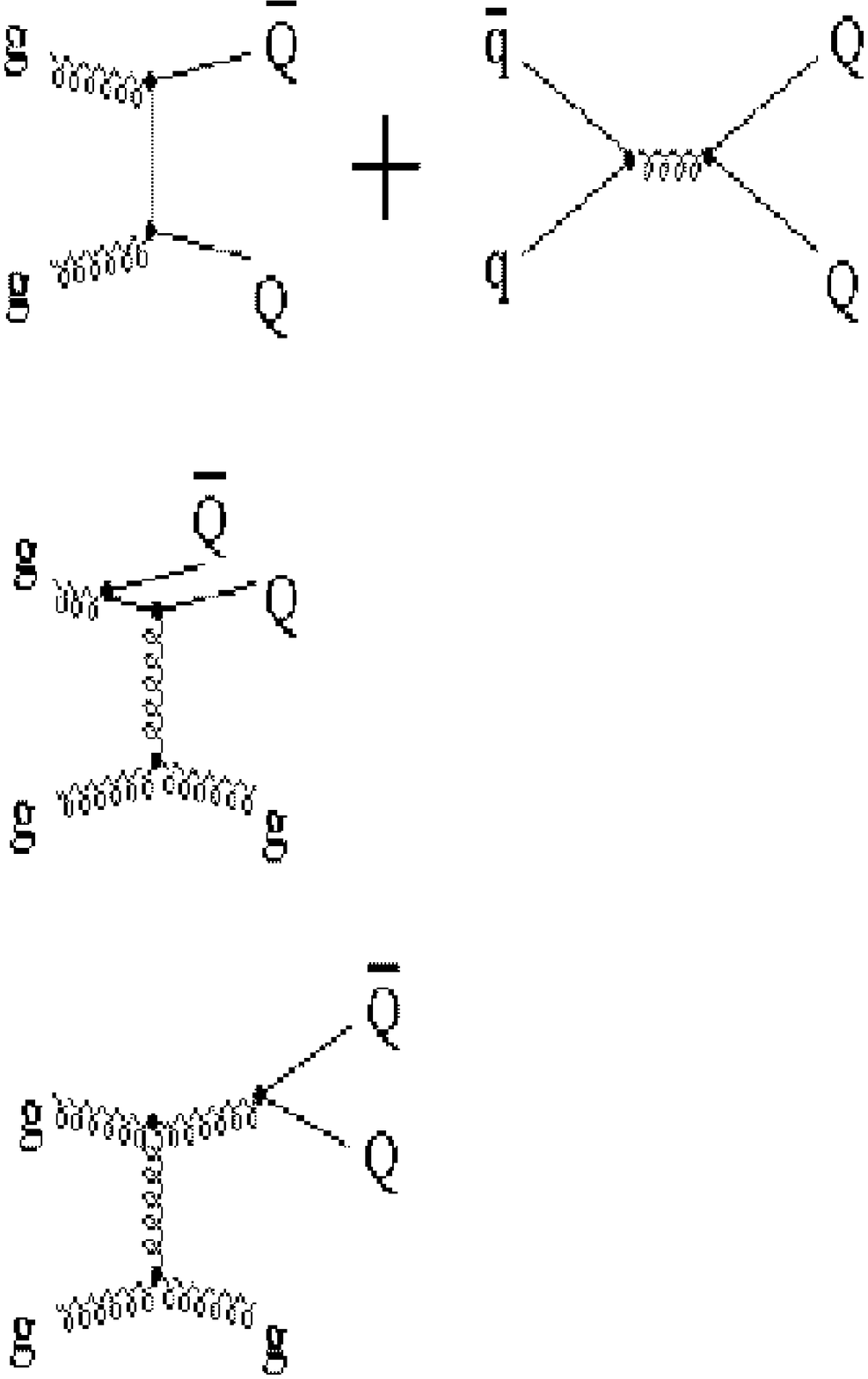}}
\caption{Examples of diagrams for different heavy quark production mechanisms 
from top to bottom correspond to the "pair creation", "flavour excitation" and
"gluon splitting".}   
\label{fig:1}
\end{minipage}
\hspace{\fill}
\begin{minipage}[t]{77mm}
\resizebox{75mm}{75mm}  
{\includegraphics{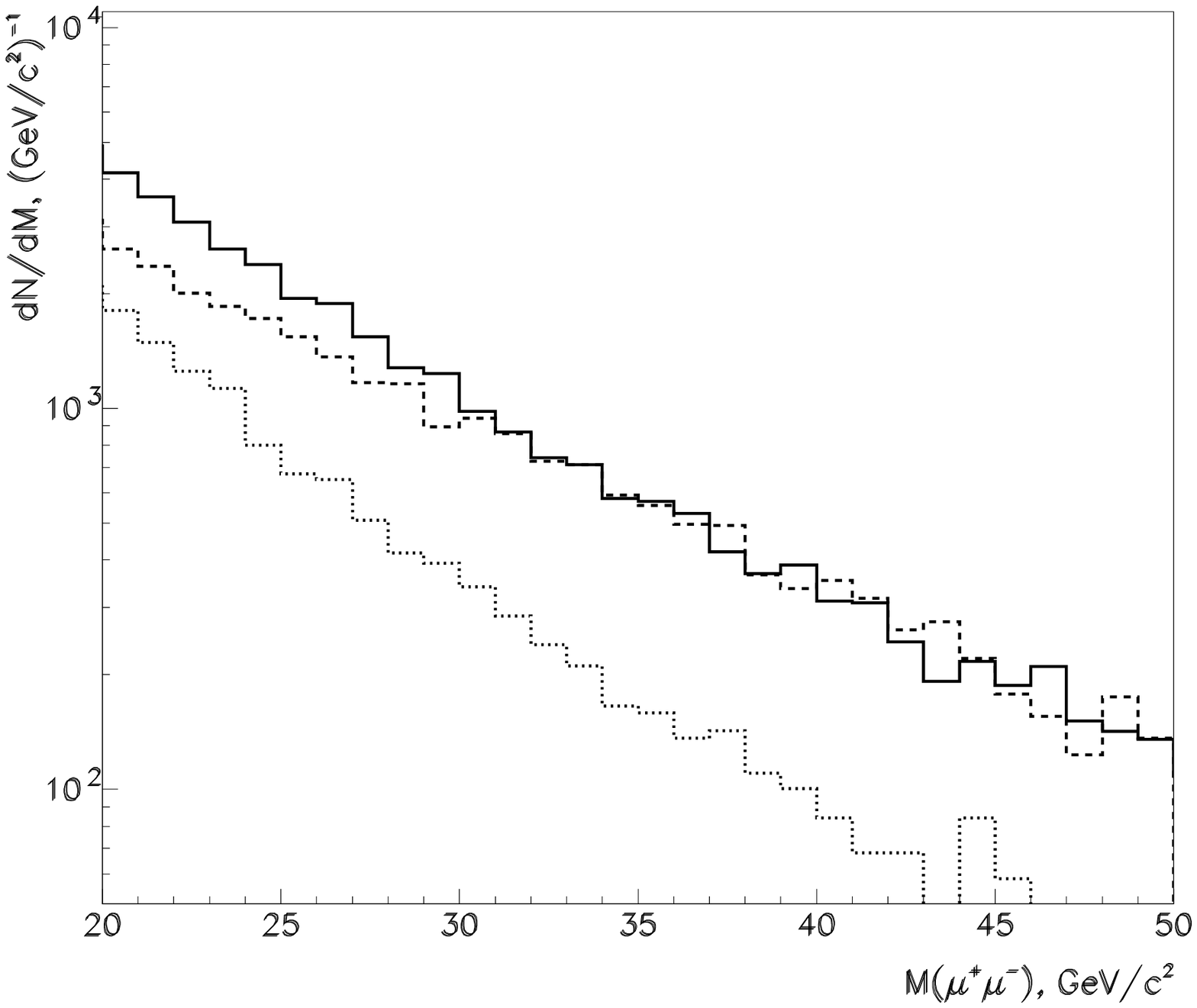}}  
\caption{Initial invariant mass distribution of $\mu^+\mu^-$ pairs from 
$B \overline{B}$ decays for different production mechanisms: 
"pair creation" (solid), "flavour excitation" (dashed) and  
"gluon splitting" (dotted).} 
\label{fig:2}
\end{minipage}
\end{figure}

In order to estimate dimuon spectra at LHC we used the initial momenta spectra, heavy
quark cross sections $\sigma _{NN}^{Q\overline{Q}}$ in $NN$ collisions and meson 
fragmentation scheme from PYTHIA$5.7$~\cite{pythia} with the default CTEQ2L parton 
distribution function (PDF) and including initial and final state radiation in vacuum 
which effectively simulates higher-order contributions to heavy quark production. 
The initial distribution of 
$Q\overline{Q}$ pairs over impact parameter $b$ of A$-$A collisions (without any
nuclear collective effects) is obtained by 
multiplying $\sigma _{NN}^{Q\overline{Q}}$ by the number of binary nucleon-nucleon 
sub-collisions: 
\begin{equation} 
\label{jet_prob}
\frac{d^2 \sigma^0_{Q\overline{Q}}}{d^2b}(b,\sqrt{s})=T_{AA} (b)
\sigma _{NN}^{Q\overline{Q}} (\sqrt{s})
\left[ 1 - \left( 1- \frac{1}  
{A^2}T_{AA}(b) \sigma^{\rm in}_{NN} (\sqrt{s}) \right) ^{A^2} \right]   
\end{equation} 
with the total inelastic non-diffractive nucleon-nucleon cross section is 
$\sigma^{\rm in}_{NN} \simeq 60$ mb at $\sqrt{s} = 5.5$ TeV. The 
standard Wood-Saxon nuclear overlap function is 
$T_{AA}(b) = \int d^2s T_A (s) T_A (|{\bf b} - {\bf s}|)$ 
where $T_A(s) = A \int  dz \rho_A(s,z) $ is the 
nuclear thickness function with nucleon density distributions $\rho_A(s,z)$. 

Figure 2 shows the initial $\mu ^+ \mu^-$ invariant high-mass spectra from dominate 
source -- semileprtonic $B\overline{B}$ decays -- in the CMS experiment~\cite{cms94} 
kinematical acceptance, $p_T^{\mu} > 5$ GeV/$c$ and $|\eta^{\mu}| < 2.4$. The total 
impact parameter integrated rates are normalized to the expected number of Pb$-$Pb 
events during a two week LHC run, $R= 1.2 \times 10^6$ s, assuming luminosity 
$L = 10^{27}~$cm$^{-2}$s$^{-1}$~\cite{cms94} to that 
$$N(\mu^+\mu^-)= R \sigma^{\mu^+\mu^-}_{AA} L .$$. 
We can see the approximately equal 
contribution from "pair creation" and "flavour excitation" at high masses, while the 
relative contribution from "gluon splitting" being decreased with $M$ increasing. The 
integrated from $M_{\mu^+\mu^-} \ge 20$ GeV/$c^2$
cross sections are $26.5$ $\mu$b ($3.2 \times 10^{4}$ events per two week), 
$20.5$ $\mu$b ($2.5 \times 10^{4}$ events) and $9.5$ $\mu$b ($1.15 \times 10^{4}$
events) for "pair creation", "flavour excitation" and "gluon splitting" respectively. 
In the remainder of the discussion we will consider as signal only dimuons from 
$b\overline{b}$ decays, because the contribution of $c\overline{c}$ fragmentation into 
kinematical region of interest is only about $1/5$ part of total rate. 
Moreover, the medium-induced charm quark energy loss can be significantly larger than 
the $b$-quark loss due to the mass difference~\cite{kharz}, resulting in an additional 
suppression of the $D\overline{D}\rightarrow  \mu^+\mu^-$ yield.

Note that main correlated background here -- Drell-Yan prompt dimuons -- is unaffected 
by medium-induced final state interactions. These dimuons are directly from the primary 
nuclear interaction vertex while the dimuons from $B$ and $D$ meson decays appear at 
secondary vertices on some distance from the primary vertex. The path length between 
the primary vertex and secondary vertices are determined by the lifetime and 
$\gamma$-factor of the mesons. This fact allows one cut to suppress the Drell-Yan rate 
by up to two orders of magnitude using the dimuon reconstruction algorithm based on the 
tracker information on secondary vertex position~\cite{lokhtin2}. 

The main uncorrelated background -- 
random hadronic decays and muon pairs of mixed origin -- appears also in the like-sign 
dimuon mass spectra and can be subtracted from the total $\mu ^+ \mu^-$ distribution 
using the $\mu ^+\mu ^+$ and $\mu ^- \mu^-$ event samples. 

Another process of particular interest is secondary $J/\psi$
production~\cite{lokhtin2}. The branching ratio $B \rightarrow J/\psi X$ is $1.15\%$.
The $J/\psi$'s subsequently decay to dimuons with a $5.9\%$ branching 
ratio. The main correlated
background here -- primary  $J/\psi$ production -- as in the case of Drell-Yan
dimuons, can be rejected using vertex detector. 
The estimated in the same kinematical region cross section for secondary $J/\psi 
(\rightarrow \mu^+\mu^-)$ production from "pair creation" is only $\approx 17\%$ 
($\approx 10.5$ $\mu$b, i.e. $1.3 \times 10^4$ events per two weeks) of total number of 
secondary  $J/\psi$'s, while the contribution from "flavour excitation" and "gluon 
splitting" being $\approx 25.5$ $\mu$b each. Since we consider here the region of much  
lower dimuon masses, $M_{\mu^+\mu^-} = M_{J/\psi} = 3.1$ GeV/$c^2$, the 
contribution of "showering" $b \overline{b}$ pairs is significantly larger as compared 
with the dimuons from semileptonic $B \overline{B}$ decays. On the 
other hand, the spectra shape of secondary  $J/\psi$'s is very similar for different 
$b \overline{b}$ production mechanisms, because it carries information about spectra of 
single $b$-quarks, which are practically the same for the different 
sources~\cite{norrbin}. The significant difference appears only for the quark-antiquark 
correlations changing variables sensitive to the $Q \overline{Q}$ 
kinematics (diquark invariant mass, azimuthal correlations). 

Note that there are theoretical uncertainties in the bottom and charm production cross 
sections in N$-$N collisions at LHC: the absolute dimuon rates depend on the choice of 
PDF, the heavy quark mass, the $B$-meson fragmentation scheme, next-to-leading order 
corrections, etc. It is therefore desirable 
that dimuon measurements in $pp$ or $dd$ collisions are made
at the same or similar energy per nucleon as in the heavy ion runs.

\section {Medium-induced energy loss of bottom quarks and dimuon spectra} 

The details of our model for the heavy quark production and passage through a 
gluon-dominated plasma, created in initial nuclear overlap zone in minimum bias
Pb$-$Pb collisions at LHC, can be found in our work~\cite{lokhtin2}. We treat the 
medium as a longitudinally expanding fluid with parton production on a hyper-surface of 
equal proper times $\tau$~\cite{bjorken} and perform a Monte-Carlo simulation of the 
mean free path $\lambda$ of heavy quark in scattering-by-scattering scheme. Then the 
basic kinetic integral equation for total energy loss 
$\Delta E$ as a function of initial energy $E$ and path length $L$ has the form 
\begin{equation} 
\Delta E (L,E) = \int\limits_0^Ldx\frac{dP(x)}{dx}
\lambda(x)\frac{dE(x,E)}{dx} ~ , ~~~~~~~~~~~~~~
\frac{dP(x)}{dx} = \frac{1}{\lambda(x)}\exp{\left( -x/\lambda(x)\right) } , 
\end{equation} 
where $x(\tau)$ is the current transverse coordinate of a quark.   

Note that although some attempts have been made to calculate medium-induced heavy quark 
energy loss for quarks of mass $M_q$(see~\cite{thoma98,zakh,kharz} for discussion), a 
full description of the coherent gluon radiation from a massive colour charge still 
lacking. There are two extreme limits for energy loss by gluon radiation.
In the low $p_T$ limit, $p_T \la M_q$, medium-induced radiation should be 
suppressed by the mass, while the ultrarelativistic limit, 
$p_T \rightarrow \infty$, corresponds to the radiation spectrum of massless quarks. 
In our case, the main contribution to high-mass dimuon and secondary charmonium  
production is due to $b$-quarks with "intermediate" values of 
$p_T \ga 5$ GeV$/c$, expected to be rather close to the incoherent regime. In order to 
estimate the sensitivity of the dimuon spectra to medium-induced effects, we consider 
two extreme cases: (i) the "minimum" effect with collisional energy loss only and (ii) 
the "maximum" effect with collisional and radiative energy loss in 
the incoherent limit of independent emissions without taking into account the LPM 
coherent suppression of radiation (i.e. $dE/dx \propto E$ and is independent of path 
length, $L$)~\cite{lokhtin2}. In the latter scenario we use the Bethe-Heitler cross 
section obtained in relativistic 
kinematics and derive the medium-induced radiative energy loss per unit 
length of quark of mass $M_q$~\cite{zakh} as the integral over the gluon radiation 
spectrum 
\begin{eqnarray}
\label{rad_los}
& &  \frac{dE}{dx} = E \rho \int \limits_0^{1-M_q/E} dy \frac{4\alpha_s 
C_3(y) (4-4y+2y^2)} {9\pi y \left[ M_q^2y^2+m_g^2(1-y)\right] } , \\ 
& & C_3 (y) = \frac{9\pi \alpha_s^2 C_{ab}}{4} \left[ 1+(1-y)^2-y^2\right]  
\ln{\frac{2\left( \alpha_s^2 \rho E y(1-y)\right) ^{1/4}}{\mu_D}} , 
\nonumber     
\end{eqnarray}  
where $m_g \sim 3T$ is the effective mass of the emitted gluon at temperature $T$; 
$y$ is the fraction of the initial quark energy carried by the emitted gluon; 
$\rho \propto T^3$ is the density of the medium; 
$C_{ab} = 9/4$, $1$ and 4/9 for $gg$, $gq$ and $qq$ scatterings respectively;    
$\alpha_s$ is the strong coupling constant for $N_f$ active quark flavours;  
the Debye screening mass $\mu_D$
regularizes the integrated parton rescattering cross section. 

The dominant contribution to the differential cross section $d\sigma / dt$ for 
scattering of a quark with energy $E$ and momentum $p = \sqrt{E^2-M_q^2}$ off 
the ``thermal" partons with energy (or effective mass) $m_0 (\tau) \sim 3T 
(\tau) \ll E$ at 
temperature $T$ can be written in the target frame as~\cite{thoma98}  
\begin{equation} 
\frac{d\sigma_{ab}}{dt} \cong C_{ab} \frac{2\pi\alpha_s^2(t)}{t^2} 
\frac{E^2}{p^2} , ~~~~~~~~~~~~ \alpha_s (t) = 
\frac{12\pi}{(33-2N_f)\ln{(t/\Lambda_{\rm QCD}^2)}},   
\end{equation} 
and QCD scale parameter $\Lambda_{\rm QCD}$ on the order of 
the critical temperature,  $\Lambda_{\rm QCD}\simeq T_c$. The integrated 
parton scattering cross section has the form: 
\begin{equation} 
\sigma_{ab}(\tau) = \int\limits_{\mu^2_D}^{t_{\rm max} }dt
\frac{d\sigma_{ab}}{dt}\>  
\end{equation} 
where $t_{\rm max}=[ s-(M_q+m_0)^2] [ s-(M_q-m_0)^2 ] / s$ and 
$s=2m_0E+m_0^2+M_q^2$. 

In the $i$-th rescattering off a comoving medium constituent (i.e. with the same
longitudinal rapidity $y$) with squared momentum transfer $t_i$ and effective
mass $m_{0i}$, the quark loses total transverse energy $\Delta e_{T i}$ and change
rapidity on $\Delta y$: 
\begin{eqnarray} 
& & \Delta e_T = E_T-\sqrt{\left( p_T-\frac{E_T}{p_T}\frac{t_i}
{2m_{0i}}-\frac{t_i}{2p_T}\right) ^2+\Delta k_t^2
\sin^2{\phi}+M_q^2} , \\  
& & \sinh {(\Delta y)} = \frac{k_t\cos{\phi}}{E_T-\Delta e_T}, 
\end{eqnarray} 
where the transverse momentum kick per scattering is 
\begin{equation} 
\Delta k_t = \sqrt
{\left( E_T-\frac{t_i}{2m_{0i}}\right) ^2-\left( 
p_T-\frac{E_T}{p_T}\frac{t_i}{2m_{0i}}-
\frac{t_i}{2p_T}\right) ^2-M_q^2} , 
\end{equation} 
and the angle $\phi$ between the direction of vector ${\bf k_{ti}}$ and 
axis $z$ being distributed uniformly. The medium-induced radiative energy loss is 
calculated with Eq.~(\ref{rad_los}) without modification of the longitudinal rapidity. 

In our calculations, we use the Bjorken scaling 
solution~\cite{bjorken} for the 
space-time evolution of the energy density, temperature and 
density of the plasma:  
\begin{equation} 
\varepsilon(\tau) \tau^{4/3} = \varepsilon_0 \tau_0^{4/3} , ~~~~~  
T(\tau) \tau^{1/3} = T_0 \tau_0^{1/3} , ~~~~~
\rho(\tau) \tau = \rho_0 \tau_0 .
\end{equation}  
To be specific, we use the initial conditions for 
a gluon-dominated plasma expected for 
central Pb+Pb collisions at LHC~\cite{esk}: 
$\tau_0 \simeq 0.1$ fm$/c$, $T_0 
\simeq 1$ GeV, $N_f \approx 0$, $\rho_g \approx 1.95T^3$. 
It is interesting that the  
initial energy density, $\varepsilon_0$, in the dense zone 
depends on $b$ very slightly,
$\delta \varepsilon_0 / \varepsilon_0 \la 10 \%$, up to $b \sim R_A$ 
and decreases rapidly for  
$b \ga R_A$~\cite{lokhtin1}. On the other hand, 
the proper time of a jet to escape the dense zone 
averaged over all possible jet production vertices, $\left< \tau_L \right> $,
is found to decrease almost linearly with increasing impact parameter. This  
means that for impact parameters $b < R_A$, 
where $\approx 60 \%$ of the heavy quark 
pairs are produced, the difference in rescattering 
intensity and the corresponding energy loss is 
determined mainly by the different
path lengths rather than the initial energy density.  

Finally, the nuclear shadowing corrections according with EKS 
model~\cite{eks} have been also taken into account in our calculations. 

The simulation of quark rescattering is halted if one of the following three
conditions is fulfilled: \\ 
1) A quark escapes from the dense zone, i.e. its path length becomes 
greater than the effective transverse spread of the matter from the production vertex 
to the escape point. The details of the geometrical calculations of these quantities 
at a given 
impact parameter can be found in Ref.~\cite{lokhtin1}. \\ 
2) The plasma cools down to $T_c=200$ MeV.  We thus neglect 
possible additional small contributions to the total energy loss due to 
re-interactions in the hadron gas. \\ 
3) A quark loses so much energy that its transverse momentum $p_T$ drops below
the average transverse momentum of the ``thermal" constituents of the medium.  
In this case, such a quark is considered to be ``thermalized" 
and its momentum in 
the rest frame of the fluid is generated from the random 
``thermal" distribution,  
$dN/d^3p \propto \exp{\left( -E/T\right) }$, boosted to the 
center-of-mass of the 
nucleus-nucleus collision~\cite{kamp98,lin98}. 

Figure 3 shows the $\mu^+\mu^-$ invariant mass spectra from $B\overline{B}$ 
decays for various nuclear effect scenarios: 
$(1)$ no loss and shadowing; $(2)$ no loss, with shadowing; $(3)$ with collisional
loss and shadowing; $(4)$ with collisional and radiative loss and shadowing. 
The absolute normalization is the same as in previous section. 
The results for all quark production mechanisms are qualitatively similar: 
the shadowing corrections are relatively small (at the level of $\approx 15\%$); 
the collisional loss reduces dimuon rate by factor $\sim 1.3-1.6$;  
the additional radiative loss reduce the rate up to factor $\sim 
3$-$4$. The relative contribution of 
radiative loss grows with increasing $M$ and $p_T$ due to the stronger energy 
dependence of the loss. The dimuon spectra from direct quarks are 
some more sensitive to the energy loss of $b$-quarks as compared with dimuons from 
"showering" quarks due to different kinematics (e.g. strong azimuthal back-to-back 
correlation). 

\begin{figure}[htb]
\begin{minipage}[t]{77mm}
\resizebox{75mm}{75mm}  
{\includegraphics{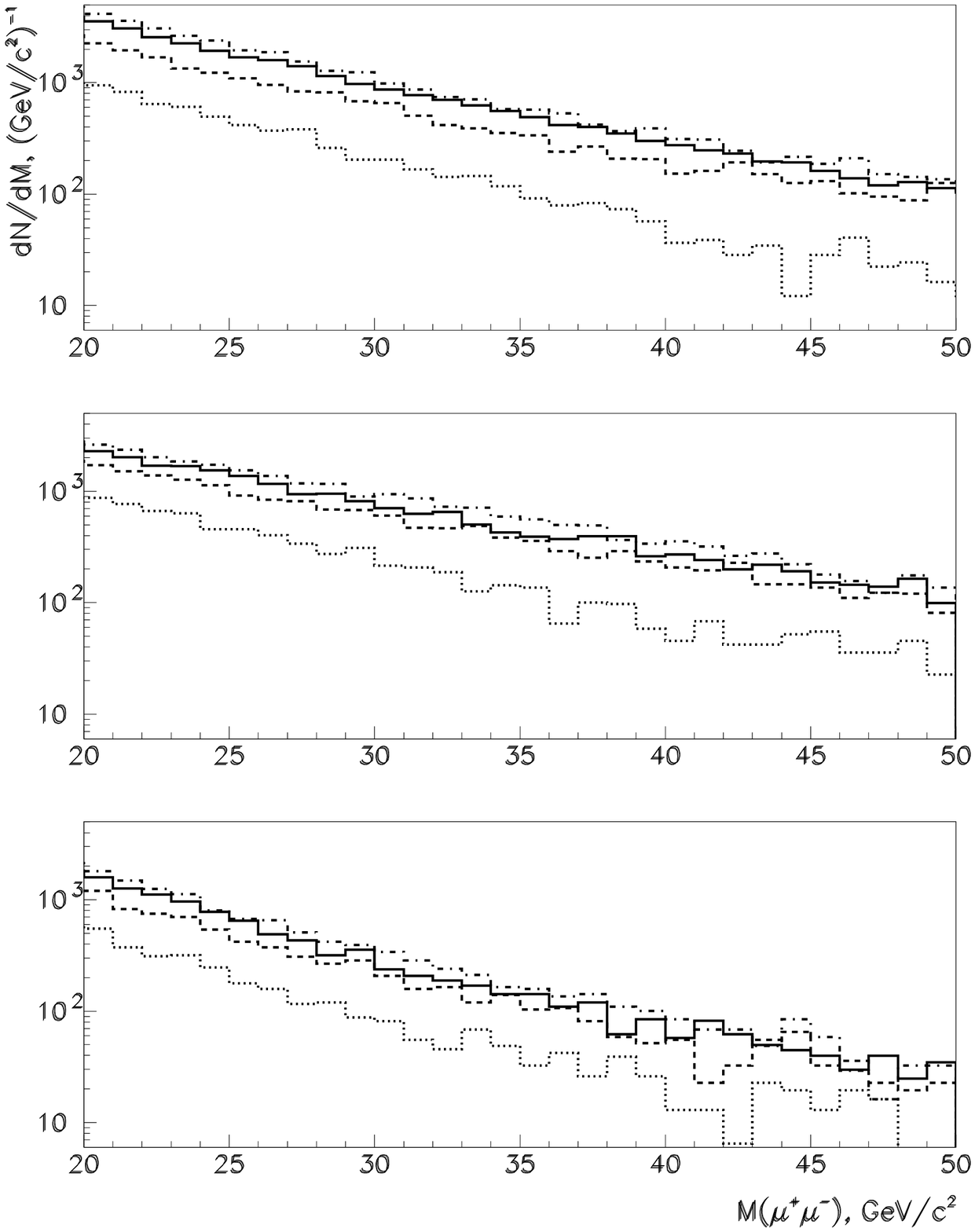}}
\caption{Invariant mass distribution of $\mu^+\mu^-$ pairs from 
$B \overline{B}$ decays for various scenarios: without loss and shadowing 
(dash-dotted), without loss and with shadowing (solid), 
with collisional loss and shadowing (dashed), with radiative and
collisional loss and shadowing (dotted). From top to bottom:  
"pair creation", "flavour excitation" and "gluon splitting".} 
\label{fig:3}
\end{minipage}
\hspace{\fill}
\begin{minipage}[t]{77mm}
\resizebox{75mm}{75mm}  
{\includegraphics{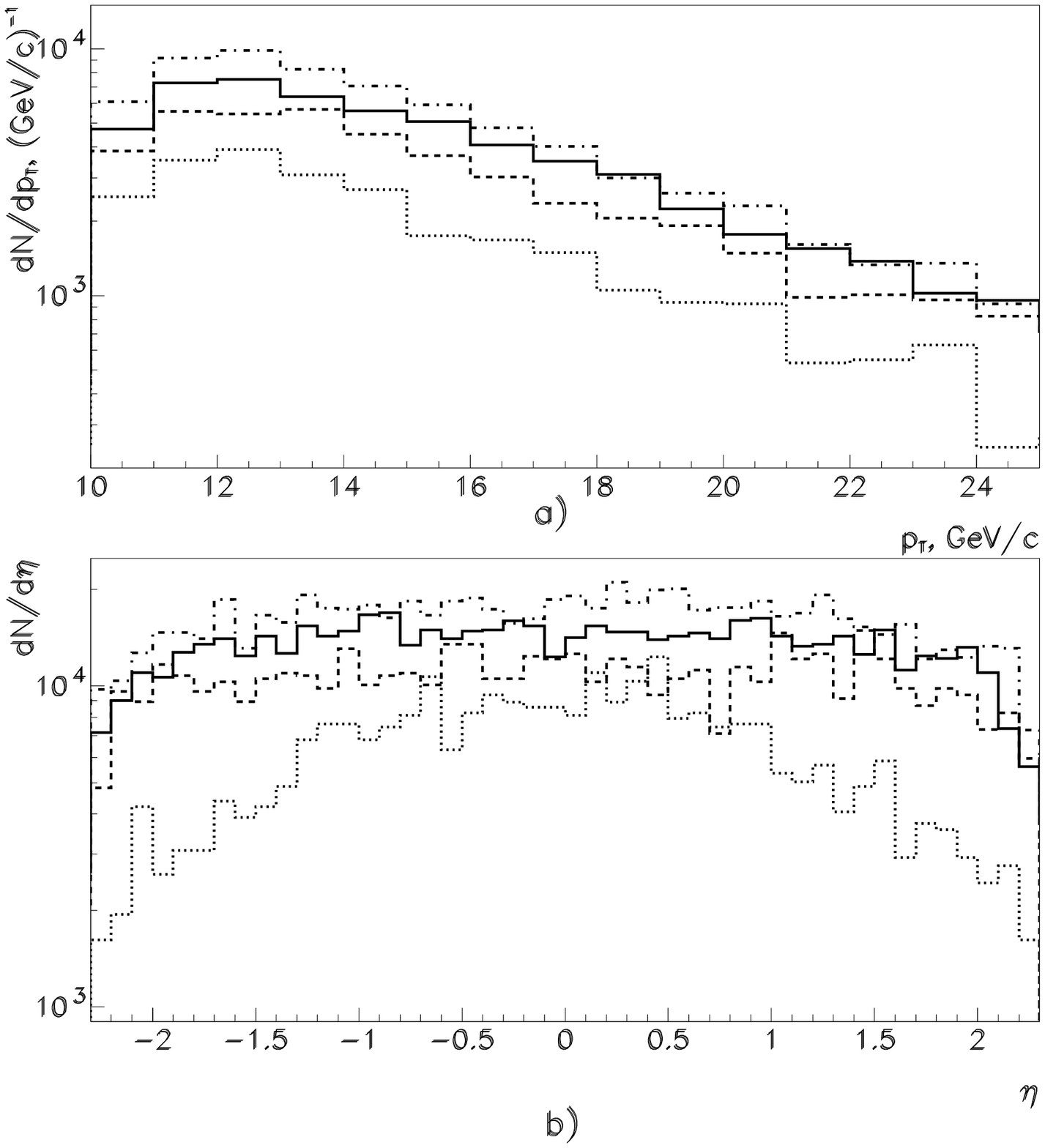}}  
\caption{Summarized over all $b \overline{b}$ production mechanisms (a) transverse 
momentum and (b)  
pseudo-rapidity distributions of $B \rightarrow J/\psi (\rightarrow \mu^+\mu^-)$
decays for various scenarios: without loss and shadowing 
(dash-dotted), without loss and with shadowing (solid), 
with collisional loss and shadowing (dashed), with radiative and
collisional loss and shadowing (dotted).} 
\label{fig:4}
\end{minipage}
\end{figure}

The summarized over all $b \overline{b}$ production mechanisms transverse momentum and 
pseudo-rapidity distributions of $B \rightarrow J/\psi (\rightarrow \mu^+\mu^-)$ decays 
is presented in figure 4. Including nuclear shadowing reduces secondary charmonium 
yield by $\sim 25\%$, while the final state rescattering 
and energy loss by $b$-quarks can further reduce the $J/\psi$ rates by a 
factor of $\sim 1.3$-$2$ in the CMS kinematical acceptance.  

We see that the influence of nuclear effects on secondary $J/\psi$'s and 
high-mass dimuons are quite different: the decrease of the  
$B \rightarrow J/\psi$ rate by nuclear shadowing is comparable to the effect of
medium-induced final state interactions. The increased sensitivity to nuclear
shadowing is due to the different $x$ and $Q^2$ regions probed. 
The different influence of energy loss on secondary charmonia and dimuons from 
$B \bar{B}$ decays is because $J/\psi$'s come from the decay of a  
single $b$ quark instead of a $b \overline{b}$ pair and there is a
non-negligible probability that the energy lost by one quark is small. 
Thus since secondary charmonia reflect energy loss of only one $b$-quark, the 
corresponding suppression is less than for semileptonic $B \overline{B}$ decays and 
is independent on quark production mechanism. Thus we believe that a comparison between 
high-mass dimuon and secondary $J/\psi$ production could clarify the nature of energy 
loss. 

\section{Conclusions} 

We have considered the different mechanisms of heavy quark production in
ultrarelativistic nuclear collisions and found that the "showering" $b \bar{b}$ pairs 
at LHC give the same order of contribution to high-mass dimuon spectra 
($M_{\mu^+\mu^-}\ge 20$ GeV/$c^2$) as direct quark pairs. The $b \bar{b}$ pair 
production from parton showers is dominant mechanism for 
$B\rightarrow J/\psi$ production. The medium-induced parton rescattering and 
collisional energy loss can reduce high-mass dimuon rate by factor $\sim 1.5$, and 
the additional radiative energy loss can reduce the rate up to factor $\sim 3$-$4$
(depending on quark production mechanism). The relative contribution of radiative loss 
grows with increasing $M$ and $p_T$ due to the stronger energy dependence of the 
loss. Due to different kinematics the high-mass dimuon spectra from direct quarks are 
more sensitive to the energy loss of $b$-quarks than dimuons from 
"showering" quarks. The nuclear shadowing corrections is $\approx 15\%$ for  
high-mass dimuons. Since secondary charmonium production reflects the 
energy loss of only one $b$-quark, the corresponding suppression by a factor of $1.3-2$ 
is lesser than for $B \overline{B}$ decays and is independent on quark production 
mechanism, but shadowing corrections are some larger.  

We conclude that the dimuon spectra will be sensitive to final state rescattering and 
energy loss of bottom quarks in dense matter. However, there are still 
theoretical uncertainties in the initial production of heavy flavours in 
nucleon-nucleon collisions at LHC energies. Thus measurements in $pp$ or 
$dd$ collisions at the same or similar energies per nucleon as in the heavy ion runs 
are required. \\ 
 
{\it Acknowledgements}. 
It is a pleasure to thank D.~Denegri and R.~Vogt for important comments. Discussions 
with M.~Bedjidian, Yu.L.~Dokshitzer, D.E.~Kharzeev, R.~Kvatadze, O.L.~Kodolova, 
L.I.~Sarycheva and U.~Wiedemann are 
gratefully acknowledged. I would like to thank the organizers of Workshop 
"Physics of the Quark-Gluon Plasma" for the warm welcome, the hospitality and creating 
a stimulating atmosphere.


\begin{thebibliography}{99}
\bibitem{ander83} B. Andersson, G. Gustafson, G. Ingelman and  T.
S\"ostrand, Phys. Rep. 97 (1983) 31. 
\bibitem{norrbin} E. Norrbin and T. Sj\"ostrand, Eur. Phys. J. C 17 (2000) 137. 
\bibitem{satz182}  H. Satz, Phys. Rep. 88 (1982) 349.  
\bibitem{muller96} J.W. Harris and B. M\"uller, Ann. Rev. Nucl. Part. Sci. 46, 
(1996) 71. 
\bibitem{lok_rev} I.P. Lokhtin, L.I. Sarycheva and 
A.M. Snigirev, Phys. Part. Nucl. 30 (1999) 279.  
\bibitem{bass_rev} S.A. Bass, M. Gyulassy, H. St\"{o}cker  
and W. Greiner, J. Phys. G25 (1999) R1.   
\bibitem{qm01} Proc. of 15th International Conference on
Ultrarelativistic Nucleus-Nucleus Collisions "Quark Matter'2001",
New York, USA, 15-20 January, 2001, in press 
\bibitem{esk} K.J. Eskola, K. Kajantie and P.V. Ruuskanen, Phys. Lett. 
B 332 (1994) 191; Eur. Phys. J. C 1 (1998) 627; 
K.J. Eskola, Prog. Theor. Phys. Suppl. 129 (1997) 1; Comments 
Nucl. Part. Phys. 22 (1998) 185; K.J. Eskola and K. Tuominen, 
Phys. Lett. B 489 (2000) 329; K.J. Eskola, K. Kajantie, P.V. 
Ruuskanen and K. Tuominen, Nucl. Phys. B 570 (2000) 379. 
\bibitem{loss_rev} R. Baier, D. Schiff and B.G.Zakharov, Annual Rev. 
Nucl. Part. Sci. 50 (2000) 37. 
\bibitem{shur97} E. Shuryak, Phys. Rev. 55 (1997) 961. 
\bibitem{kamp98} B. Kampfer, O.P. Pavlenko and K. Gallmeister, 
Phys. Lett. B 419 (1998) 412.  
\bibitem{lin98} Z. Lin, R. Vogt and X.-N. Wang, Phys. Rev. C 57 (1998) 899; 
Z. Lin and R. Vogt, Nucl. Phys. B 544 (1999) 339.  
\bibitem{lokhtin2} I.P. Lokhtin and A.M. Snigirev, Eur. Phys. J. C 21 (2001) 155. 
\bibitem{kharz} Yu.L. Dokshitzer and D.E. Kharzeev, e-print hep-ph/0106202; 
D.E. Kharzeev, contribution in Workshop "Physics of the Quark-Gluon Plasma"
\bibitem{pythia} T. Sj\"ostrand. Comput. Phys. Commun. 82 (1994) 74.  
\bibitem{cms94} CMS Collaboration Technical Proposal, CERN/LHCC 94-38, 1994. 
\bibitem{bjorken} J.D. Bjorken, Phys. Rev. D27 (1983) 140.  
\bibitem{thoma98} M.G. Mustafa, D. Pal, D.K. Srivastava and 
M. Thoma, Phys. Lett. B428 (1998) 234.  
\bibitem{zakh} B.G. Zakharov, JETP Lett. 65 (1997) 615; Phys. At. Nucl. 61 (1998) 838.  
\bibitem{lokhtin1} I.P. Lokhtin and A.M. Snigirev, Eur. Phys. J. C 16 (2000) 527.
\bibitem{eks} K.J. Eskola, V.J. Kolhinen and C.A. Salgado, Eur. 
Phys. J. C 9 (1999) 61.  
\end{thebibliography}
\end{document}